\documentclass[pra,twocolumn,aps,epsf,showpacs,superscriptaddress,longbibliography]{revtex4-2}

\usepackage[dvipsnames]{xcolor}
\usepackage{physics}
\usepackage{amsmath}
\usepackage{amsfonts}
\usepackage{dsfont}

\newcommand{\jk}[1]{{#1}}

\usepackage{graphicx}
\begin{document}
	
	\title{Stability of topologically protected slow light against disorder}
	
	\author{Jonas F.~Karcher}
	\affiliation{{Pennsylvania State University, Department of Physics, University Park, Pennsylvania 16802, USA}}
	
	\author{Sarang Gopalakrishnan}
	\affiliation{{Princeton University, Department of Electrical and Computer Engineering, Princeton, NJ 08544, USA}}
	
	\author{Mikael C.~Rechtsman}
	\affiliation{{Pennsylvania State University, Department of Physics, University Park, Pennsylvania 16802, USA}}

	\begin{abstract}
		Slowing down light in on-chip photonic devices strongly enhances light-matter interaction, but typically also leads to increased backscattering and small-bandwidth operation. It was shown recently~\cite{Guglielmon2019} that, if one modifies the edge termination of a photonic Chern insulator such that the edge mode wraps many times around the Brillouin zone, light can be slowed to arbitrarily low group velocity over a large bandwidth, without being subject to backscattering.  Here we study the robustness of these in-gap slow light modes against fabrication disorder, finding that disorder on scales significantly larger than the minigaps between edge bands is tolerable.  We identify the mechanism for wavepacket breakup as disorder-induced velocity renormalization and calculate the associated breakup time. 
	\end{abstract}

	\maketitle

	\section{Introduction}
	Slow light has been of great interest due to its potential applications in a variety of fields, including classical~\cite{corcoran2009green} and quantum~\cite{hau1999light} nonlinear optics, quantum information processing~\cite{chang14}, optical communications~\cite{zhao2009}, and sensing~\cite{shahriar2007}. Slowing down light on a photonic chip is typically achieved through the use of photonic crystal waveguides that exhibit photonic bandgaps~\cite{krauss2007slow,vlasov2005active,baba08}, where there is a reduction in the group velocity of light at the band edge. This has the effect of making light interact more strongly with matter, potentially leading to significant enhancement of nonlinear processes such as frequency comb~\cite{vasco2019} and entangled pair~\cite{chang14} generation. 
	
	Slow light waveguides suffer from significant shortcomings, however.  One is narrow bandwidth: by construction, light is only slow in a small range of frequencies near the band edge, where the group velocity goes to zero. Another is that slow light wavepackets are susceptible to losses and fluctuations due to fabrication imperfections. Backscattering is strongly enhanced due to the slow velocity (and accompanying high density of states) and thus slow-light waveguides are prone to Anderson localization~\cite{melati14}, limiting their utility.  While out-of-plane radiation loss is enhanced as the group velocity decreases, backscattering is the greater obstacle near the band edge (these forms of loss scale as $1/v$ and $1/v^2$, respectively, where $v$ is the group velocity)~\cite{hughes2005extrinsic, rosiek2023observation}.
	
	Photonic crystal Chern insulators are materials that exhibit chiral edge states which are resistant to backscattering. These edge states can be implemented in various systems such as magnetooptical photonic crystals~\cite{wang09} and non-planar geometries such as waveguide arrays~\cite{rechtsman2013photonic}.  In another approach, reciprocal topological ring resonator arrays were proposed as delay lines~\cite{hafezi2011robust, hafezi2013imaging}.  Initial attempts have been made to realize topological magneto-optical photonic crystals in optics~\cite{bahari2017nonreciprocal}, and Floquet Chern insulators of light have been demonstrated in nonlinear photonic crystals~\cite{he2019floquet,jin2023observation}, but it remains a materials challenge to enlarge the topological band gap.  In all these cases, chiral edge modes traverse the entire bulk band gap within a single Brillouin zone. 
	
	A recent proposal by Guglielmon et al.~\cite{Guglielmon2019} suggests that modifying the edge termination of a photonic Chern insulator can simultaneously address the fundamental obstacles of narrow-band operation and parasitic backscattering by disorder. Specifically, the proposed modification involves winding the topological edge state multiple times around the Brillouin zone, as illustrated in Figure \ref{fig:ren1}. This winding allows for robust slow light propagation over a wide range of frequencies without the possibility of Anderson localization.  However, it has remained unclear how much disorder can be added to such Chern insulator slow light systems before the effect breaks down.  Typically, topological observables (such as the Hall conductance, for example) remain robust until disorder causes the bulk band gap to close; here, however, it is reasonable to guess that disorder must be smaller than the `minigap' between edge bands, which may be significantly less than the whole gap, depending on the number of windings around the Brillouin zone.  Here, by analytically and numerically examining wavepacket break-up time, we show that for reasonably-sized devices, the amount of disorder can be significantly larger than the minigap between edge bands.  We show analytically that wavepacket break-up originates from disorder-induced group velocity renormalization of the edge states.  Furthermore, for a given device size and number of edge windings, we provide a recipe for quantifying the amount of disorder that is tolerable to maintain device operation.

	\section{Models and methods}
	In this sections, we present the different complementary models considered in this work. Then we define a condition for the breakup of wavepacket transport.
	\subsection{Models}
	\label{sec:models}
\begin{figure}
		\centering
		\includegraphics[width=0.45\textwidth]{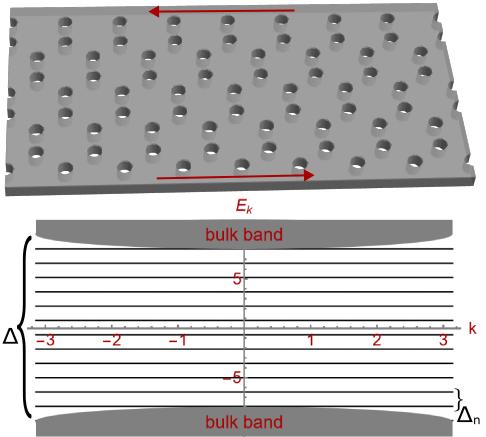}
		\caption{{\bf Schematic diagram of a photonic crystal slab}, which may form a Chern insulator if it is non-reciprocal, implying the presence of backscatter-free topological edge modes. Right and left movers (arrows) are spatially separated by the bulk. Edge tapering as in Ref.~\cite{Guglielmon2019} allows for a non-dispersive wrapping of the edge state around the edge Brillouin zone several times , here $n=6$.  Only right-movers are shown in the dispersion diagram.  }
		\label{fig:ren1}
	\end{figure}

	Figure~\ref{fig:ren1} shows a schematic diagram of a two-dimensional photonic crystal Chern insulator slab, which can be realized using a magneto-optical material~\cite{wang09, skirlo2015experimental}, or by nonlinear temporal modulation \cite{jin2023observation}.  Furthermore, it has been demonstrated that the multiple winding procedure of Ref.~\cite{Guglielmon2019} can be implemented in a magneto-optical photonic crystal model~\cite{mann2021}; a depiction of the associated band structure is shown in Fig.~\ref{fig:ren1}.  We show below that in the presence of disorder, the velocity of the edge state gets renormalized, and this effect is the leading cause of wavepacket breakup on the topological slow light edge.
	
	Following Refs. \cite{Stacey_1982, Chou2014chalker}, we can write down an effective theory for the topological edge state Hamiltonian in momentum space (this equation defines a linear dispersion in k-space, with the possibility of coupling between different $k$ via disorder):
	\begin{align}
		(H_{\rm edge})_{k,k'} &= v_e \hbar k \delta(k-k') + V_{k,k'},
		\label{eq:ham_low}
	\end{align}
	where $v_e$ is the uniform \jk{group} velocity in the clean model. Here $k$ lives in the extended zone scheme, i.e. wraps $(2n+1)$ times around the Brillouin zone $k\in(-\pi/a,\pi/a)$. Due to this edge wrapping, the velocity $v$ asymptotically goes as \jk{$(2n+1)^{-1}\Delta a/2\pi$} for many channels, where $\Delta$ is the band gap. In other words, the energy wrapping occurs over the size of the subband gap $\Delta_n\sim (2n+1)^{-1}\Delta$. Since the left mover is spatially separated from the right movers by the bulk, we can safely neglect coupling between them even in the presence of disorder $V(x)$, since all \jk{state overlaps} will be exponentially suppressed with the bulk width.
	In the Appendix~\ref{sec:mod_imp}, we motivate that intersubband scattering is negligible by studying \jk{state overlaps} in the full bulk Hamiltonian.  Indeed, for a qualitative understanding of the physics, the edge theory shown above is sufficient.  As we show in Fig.~\ref{fig:imp_matrix2}, there is good agreement between the continuum Hamiltonian of Eq.~\eqref{eq:ham_low} and the full lattice model of a Chern insulator.
	
	\subsection{Definition of breakup time and length}
	To determine the breakup time $\tau$ (and therefore maximal length of the device $L_{\rm max}=v \tau$), we use the criterion that the accumulated spatial decoherence due to disorder-induced dispersion  
	is of the order of the size of the wavepacket $\Delta l$ (which is inversely proportional to the energy bandwidth $\delta \epsilon$):
	\begin{align}
		(\delta \epsilon)^{-1} \sim \Delta l  \sim (v(\epsilon+\delta \epsilon) - v(\epsilon)) \tau \sim  v'(\epsilon)\delta \epsilon \tau ,
	\end{align}
	where $\delta\epsilon$ is the energy bandwidth and $v(\epsilon)$ is the group velocity of the edge state as a function of energy, $\epsilon$. \jk{If this linearization of the velocity is not valid because the dispersion is not sufficiently optimized, then there is breakup already without the effect of fabrication disorder. For our scaling analysis we do not consider this case. }
	Using this, we can solve for the minimal characteristic breakup time:
	\begin{align}
		\tau \sim \dfrac{1}{(\delta \epsilon)^2\max_{|\omega|<\delta\epsilon}|v'(\epsilon+\omega)| } \label{eq:prop}.
	\end{align}
	This is nothing more than the characteristic dispersion time associated with the effective band structure as renormalized by the disorder \jk{, this is an effect most prominent in Dirac systems \cite{wilson2020disorder, wang2013velocity, chae2012renormalization}. The topological protection here is crucial for the renormalization to matter: in non-topological 1D systems with slow group velocity, localization sets in at weak disorder. Due to the localization of eigenstates the group velocity ceases to well defined. The chirality of the edge states prevents them from localizing and keeps them plane wave-like (this is the ballistic regime \cite{Evers2008}: momentum is approximately a good quantum number and the (renormalized) group velocity matters). In that sense, systems with chiral edge states stay in the ballistic regime for arbitrary disorder (unless the bulk gap closes).} 
	
	Using this scaling relation \eqref{eq:prop}, we can now establish an estimate for the allowed device size (parallel to the edge) such that the wavepacket will stay spatially coherent and will thus be useful for linear and nonlinear optical applications. This depends on the energy bandwidth of the wavepacket $\delta \epsilon$ and the energy dependence of the renormalized velocity $v(\epsilon)$. Below, we (i) use weak disorder perturbation theory to show the emergence of an energy dependent velocity and (ii) do an exact diagonalization study of the full bulk Chern insulator Hamiltonian to validate our analytical predictions and to go beyond perturbation theory. 
	
	\begin{figure}
		\centering
		\includegraphics[width=.99\linewidth]{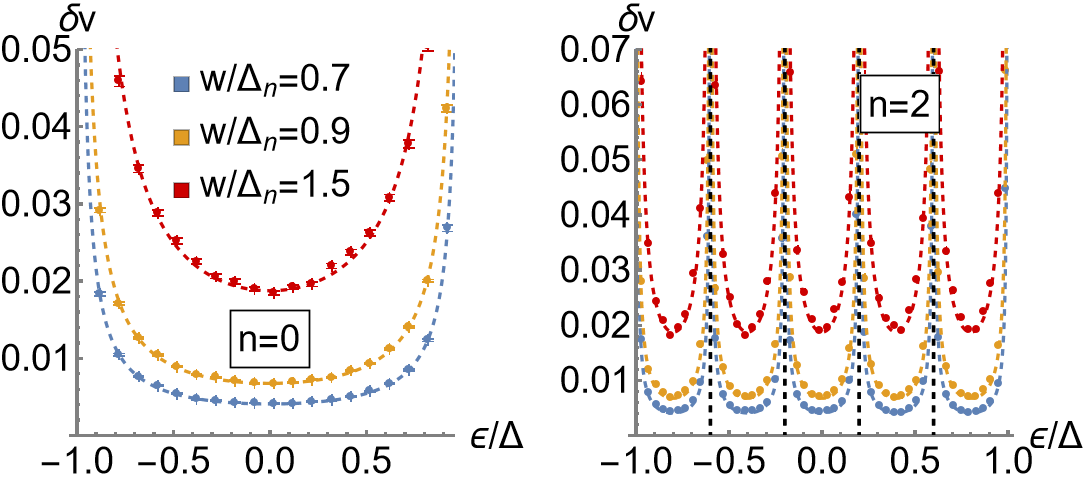}
		\caption{{\bf Velocity renormalization:} Velocity renormalization, $\delta v$ for (a) $n=0$ and (b) $n=2$.  The \jk{dashed lines} are the analytical result from Eq.~\eqref{eq:vel_ren_ana}, and the \jk{data points} are numerical time evolution of the continuum edge model (as described in Sec. \ref{sec:mom_break} of the appendix).  The system size is $L=2500$, \jk{disorder strength $0.7\Delta_n \leq w \leq 1.5\Delta_n$} and $\delta \epsilon = \Delta_n/10$. \jk{Vertical black} dashed line highlights subband boundaries. } 
		\label{fig:ren5}
	\end{figure}

	\section{Velocity renormalization: continuum vs. lattice}
	We first provide a perturbative argument for how the chiral edge state velocity is renormalized because of the finite bandwidth $\Delta = 2\pi v/a$ of the edge modes on the lattice. (Here, the continuum limit is $\Delta \to \infty, a \to 0$; in this limit, the system is known to be stable to arbitrarily strong potential disorder \cite{Xie2016}.) For simplicity we begin with the $n = 0$ case, with a single sub-band. The disorder \jk{enables} a uniform scattering \jk{process} between any two edge momenta $k, k'$. Since backscattering is absent, the main effect of the scattering is to shift the energy of each mode according to second-order perturbation theory. The modes repel each other: a given mode is pushed up (down) by modes lower (higher) in energy. The net shift of the mode is given by the imbalance between these effects, and therefore increases as one moves away from the center of the sub-band. Since modes that are farther from the center of the sub-band receive larger shifts, disorder makes the sub-band steeper, so the renormalized velocity exceeds the bare velocity. This effect relies on the edge modes living in a finite energy window, so it vanishes in the continuum limit $\Delta \to \infty$. In the $n > 0$ case, the scattering processes of the disorder among different sub-bands can be neglected, as we show in Appendix~\ref{sec:mod_imp} by analyzing the eigenstate overlap for different subbands. We find that the intraband overlap clearly dominates over the interband overlap. Consequently, this velocity-dependent renormalization takes place separately for each sub-band, leading to the periodic modulation of velocity seen in Fig.~\ref{fig:ren2}.
	
	To sharpen this intuitive argument we compute the 
	perturbative corrections to the self energy, $\Sigma_k(\omega)$, in the weak disorder limit.
	A technical calculation in Sec.~\ref{sec:self} of the appendix yields
	the lowest order result for the velocity renormalization for $n=0$ in terms of the clean velocity $v_e$ and disorder strength $w$ (assuming $V(x) $ to be drawn from $[-w/2,w/2]$ independently):
	\begin{align}
		\delta v(\epsilon) = v_e\dfrac{w^2/12\pi}{\Delta^2-\epsilon^2}.\label{eq:vel_ren_ana}
	\end{align}
	If bulk bands are neglected, disorder therefore always increases the effective velocity. When bulk bands are considered, there can also be a reduced effective velocity. Now that we have perturbatively derived the change in velocity as a result of the presence of disorder, we have established, using Eq.~\eqref{eq:prop}, that there is indeed a finite breakup time. This analysis is well controlled for weak disorder but numerical results show that it holds accurate up to approximately the scale of the mini-gap, $\Delta_n$.  This approach yields an intuition for the effect of disorder on the edge bands with multiple winding: disorder acts to undo the slow group velocity of the edge state, effectively speeding it up.  In Fig.~\ref{fig:ren5}, we plot both the change in group velocity due to disorder, $\delta v$, as a function of disorder, $w$, as well as the breakup time that results from the dispersion that results.  Below, we show that these effects persist in numerical results in the low energy model as well as the full bulk model of a Chern insulator with a multiply-winding edge state.
	
	\section{Comparison to full model of Chern insulator on a lattice}
	Above, we used a continuum description of the multiply-winded chiral edge state in order to calculate the effect of disorder on the velocity renormalization, and hence the wavepacket breakup time.  Here, we use second-order perturbation theory to compute the velocity renormalization using a full lattice model of a Chern insulator, for small disorder $w\ll \Delta_n$, in order to compare with numerical results.  According to second-order perturbation theory, the eigenenergies are:
	\begin{align}
		E_k^{(2)} &= E_k + \langle k| V|k\rangle + \sum_{k\neq k'}\dfrac{|\langle k| V|k'\rangle|^2}{E_{k'}-E_k}
	\end{align}
	for a fixed disorder configuration $V$. 
	We assume white noise correlations $\langle V(x) V(x')\rangle = \frac{1}{12}w^2\delta(x-x') $ and an unbiased potential $\langle V(x)\rangle =0 $. From this, we can obtain the renormalized group velocity as a momentum derivative of the average dispersion  $v^{(2)}=\partial_k\langle E_k^{(2)} \rangle$:  
	\begin{align}
		v^{(2)} &= v_e + a_v w^2, &a_v &= \partial_k\sum_{k'\neq k}\sum_x\dfrac{|\langle k| x\rangle  \langle x|k'\rangle|^2}{E_{k'}-E_k}  \label{eq:vel_ren}.
	\end{align}
	We will then compare this directly to our results in the continuum model discussed above.

	\begin{figure}
		\centering
		\includegraphics[width=0.99\linewidth]{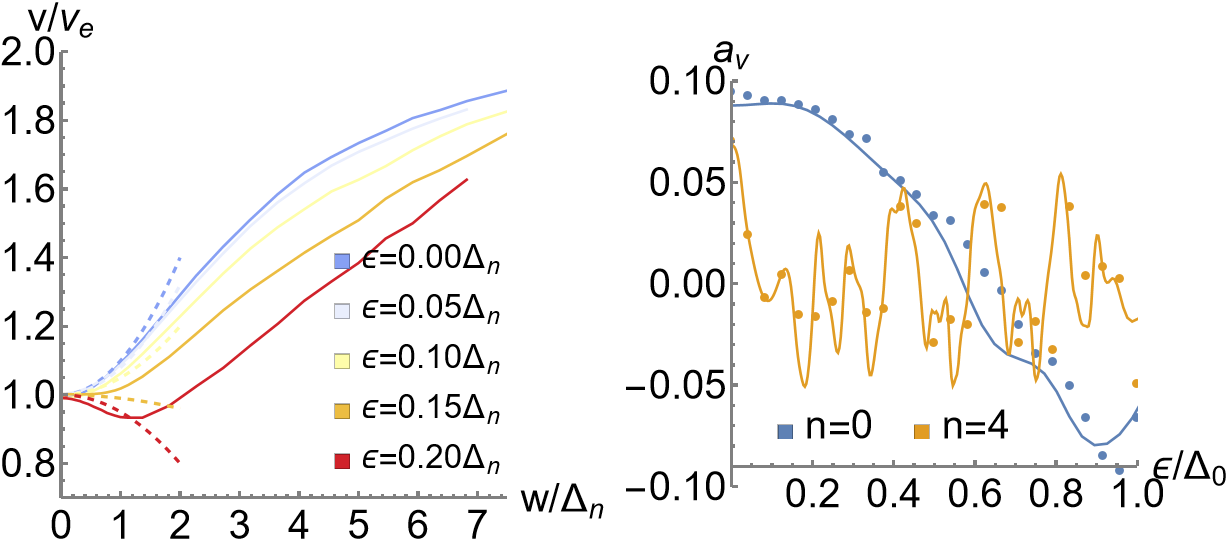}
		\caption{\jk{{\bf Velocity renormalization as a function of disorder and winding number:} (a) Velocity renormalization $v(w,\epsilon)/v_e$ for $n=0$ winding as function of disorder strength $w$. The different colors correspond to different energies, $\epsilon$. At weak disorder ($w\lesssim\Delta_n$), the renormalization is quadratic in $w$ with an energy-dependent coefficient $a_v$.  The coefficient matches weak-disorder perturbation theory as calculated using Eq.~\eqref{eq:vel_ren} well into the intermediate disorder regime (dashed lines). (b) Quadratic coefficient of the dependence of velocity renormalization, $\delta v$, on disorder $w$ for winding numbers $n=0,4$.  The solid lines are calculated from Eq.~\eqref{eq:vel_ren} and the dots from full numerical time evolution.}
		}
		\label{fig:ren2}
	\end{figure}

	The specific model we study is the Chern insulator from Ref.~\cite{Guglielmon2019} with multiple edge state winding, for different winding numbers $n$. The system size in the direction parallel to the edge is $L=2500$ unit cells and we choose the transverse direction to be $W=50$ unit cells in order to minimize the overlap between edges. Any overlap would render the system trivial by coupling the edge states on opposite sides, leading to backscattering.

	To numerically calculate the velocity renormalization, we examine the time evolution of an energetically narrow wavepacket overlapping roughly $100$ consecutive eigenstates. We track the position $x(t)$ and width of the wavepacket $\sigma_x(t)$ in real space during time evolution. In Fig.~\ref{fig:ren2} we show the velocity renormalization $v(w,\epsilon)/v_e$ for $n=0$ winding as function of disorder strength $w$. At weak disorder (smaller than the reduced gap $\Delta_n$), the renormalization is quadratic in $w$ with an energy dependent coefficient.  This numerical result matches the analytical prediction of Eq.~\eqref{eq:vel_ren}. 
	An important point is the reduction of group velocity that occurs in the vicinity of the bulk bands: for an evenly spaced edge state spectrum the disorder-induced energy level repulsion always leads to an increase of the group velocity. In the bulk model, the evenly spaced chiral edge states bleed into the bulk bands. The dense bulk bands repel the edge states much more strongly and compress the energy range occupied by the chiral edge states, leading to a reduction of their group velocity near the band edge.
	We investigate the dependence on winding number in Fig.~\ref{fig:ren2}, where we show the results from Eq.~\eqref{eq:vel_ren} for different $n$ compared to the numerical data from wave packet tracking (shown in Fig.~\ref{fig:ren2a} of Appendix~\ref{sec:cprbm}). For $n=0$, the results match well. At large winding $n=4$, there is still a fair agreement, despite the rapid oscillations of the velocity renormalization $\delta v(\epsilon)/v_e$ as function of energy. The deviations can be explained by the fact that a wavepacket always has finite support over a range of energy bandwidth $\delta \epsilon$. \jk{Consequently the fastest oscillations of the renormalized velocity $\delta v(\epsilon)$ as functions of energy $\epsilon$ are smeared out on the scale of the energy bandwidth of the wavepacket $\delta \epsilon$.}

	\begin{figure}
		\centering
		\includegraphics[width=0.45\textwidth]{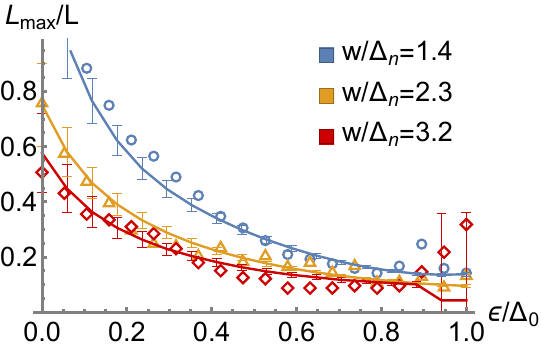}
		\caption{{\bf Chern insulator breakup simulations.} Breakup length $L_{\rm max}(w,\epsilon)$ normalized by the system size $L=2500$ for $n=0$ winding as function of energy $\epsilon$ \jk{and disorder strength $w$}. The energy bandwidth $\delta \epsilon = \Delta_n/4$. Both the disorder and the size of the wavepacket are chosen to be relatively large in order to observe the breakup cleanly. The dots represent full numerical results and the solid line is the analytical result of Eq.~\eqref{eq:lmax_weak}. This shows that the wavepacket breaks up more quickly away from the center of the gap. }  
		\label{fig:ren3}
	\end{figure}
	
	We can assign a breakup time $\tau$ (and correspondingly length $L_{\rm max} = v(\epsilon) \tau$) in terms of dispersive broadening beyond a threshold $\sigma_x(t=\tau)=C\sigma_x(t=0)$, where we choose $C=4.5$ for our numerical calculations throughout. In Fig.~\ref{fig:ren3} we show the breakup length $L_{\rm max}(w,\epsilon)$ as a function of energy for $n=0$ winding. 
	
	The solid line is the inverse derivative of the renormalized velocity.  Near the band center ($\epsilon$ much smaller than the subband gap $\Delta_n$), the breakup time becomes very large and is hard to determine numerically. For realistic wavepackets more narrow in energy $\delta \epsilon \ll \Delta_n$, breakup occurs only for much larger propagation distances even near the band edge.

	\section{Scaling form for breakup length}
	Given the characteristic energy scale $\Delta_n$ over which $v(\epsilon)$ varies, we can estimate the derivative $v'(\epsilon)\sim \delta v_e / \Delta_n \sim (v_e/\Delta_n) \cdot (w^2/\Delta_n^2)$. Consequently the maximum system size scales like
	\begin{align}
		L_{\rm max} = v\tau \sim \dfrac{v}{v'(\epsilon) (\delta \epsilon)^2} \sim \dfrac{\Delta_n^3}{ (\delta \epsilon)^2 w^2} \label{eq:lmax_weak}.
	\end{align}
	For very narrow energy bandwidths, wavepackets are not strongly localized in real space $\delta \epsilon \ll \Delta_n$, and thus the maximum device size before breakup $L_{\rm max}$ can be quite large for up to moderate disorder $w\gtrsim \Delta_n$.  
	
	In the regime of strong disorder (in Sec.~\ref{sec:robust} of the appendix it is shown that the engineered edge states are robust to strong disorder even for relatively small widths $W$), the eigenstates are no longer close to plane waves and the velocity renormalization is strong. Breakup is driven by both effects in this case.
	
	\subsection{Comparison to non-topological case} The localization length in a conventional one-dimensional waveguide (dispersion $\epsilon_k \sim \cos(ka)$, where $k$ is the wavenumber and $a$ is the lattice constant) can be written as:
	\begin{align*}
		\xi(\epsilon) &= 24 v(\epsilon)^2 w^{-2},
	\end{align*}
	in terms of the velocity. This is an upper bound on break-up length (strong localization sets in at this scale). Our device setup allows for much larger fabrication disorder to be present for narrow-band wavepackets. In other words, the smallness of the energy bandwidth $\delta \epsilon$ increases our maximum size quadratically (see Eq.~\eqref{eq:lmax_weak}).  Furthermore, in practical photonic crystal devices, achieving slow light requires operating very close to the band edge, severely limiting the operational bandwidth.  By contrast, the only limitation on the bandwidth in the Chern insulator is the size of the bulk band gap. 
	
	\subsection{Application to realistic device parameters} To illustrate how the formula for the breakup length (Eq.~\eqref{eq:lmax_weak}) is applied to a realistic photonic Chern insulator, consider a photonic crystal edge waveguide with lattice constant $a=500\mathrm{nm}$, bulk gap at $\sim 1500nm$ of $10\;{\rm nm}$, $n=20$ windings and energy bandwidth $\hbar \jk{\delta  \epsilon}^{-1}\sim \tau_{\rm pulse} = 10\;{\rm ps}$, this gives a maximum device length
	\begin{align*}
		L_{\rm max} \sim 2\pi a \dfrac{(\Delta_n)^2 (\Delta_0)^2}{w^2 (\jk{\delta  \epsilon})^2} = 100\;{\rm \mu m}
	\end{align*}
	for $w\sim \Delta_n$.  This formula therefore gives a prescription for the design of the slow-light waveguide: for a known amount of known fabrication-induced disorder on an integrated photonic chip, and a given device size, the maximal number of windings of the edge state can be calculated.  This gives the slowest possible group velocity that can be achieved in the device, while maintaining the spatial coherence of the pulse.

	\section{Discussion} Our key result is that for a photonic Chern insulator system with a multiple-winding edge state, wave packet breakup is caused by disorder-induced velocity renormalization, or dispersion.  The effective dispersion leads to the breakup (or spatial decoherence) of the wavepacket, giving a typical length over which the edge can act as a useful slow light waveguide. We have numerically verified that this formula is valid for systems of sizes typical for experimental realizations and disorder strengths $w\sim 7 \Delta_n$ of several times the subband gap.  A future research direction is the investigation of the effects of nonlinearity in slow light chiral edge state waveguides, specifically frequency comb generation and the generation of entangled photons by spontaneous down conversion and four-wave mixing.  If a sufficiently large topological band gap can be opened in a two-dimensional photonic crystal (whether by magneto-optics \cite{wang09, haldane2008possible}, or temporal modulation \cite{jin2023observation}), we expect that the multiple-winding disordered chiral edge state described here can overcome the problem of Anderson localization and allow for much slower group velocities and thus greater efficiency of slow-light waveguides.       
	
	\vspace{.4cm}
	\section*{Acknowledgement} 
	We thank M. J$\ddot {\text {u}}$rgensen for useful discussions. We acknowledge support by the Army Research Office under the MURI program, grant number W911NF-22-2-0103. S.G. acknowledges support from NSF DMR-2334056.
	
	\bibliography{Quasi1D}

	\appendix 
	\onecolumngrid
	
	\section{Higher winding numbers $n>0$}
	\label{sec:cprbm}
	In the main text, we mainly considered the zero winding case, since it is computationally most easily tractable. Here, we show the detailed results supporting our claim of generality of the breakup mechanism for arbitrary winding numbers.
	
\begin{figure*}
		\centering
		\includegraphics[width=0.9\textwidth]{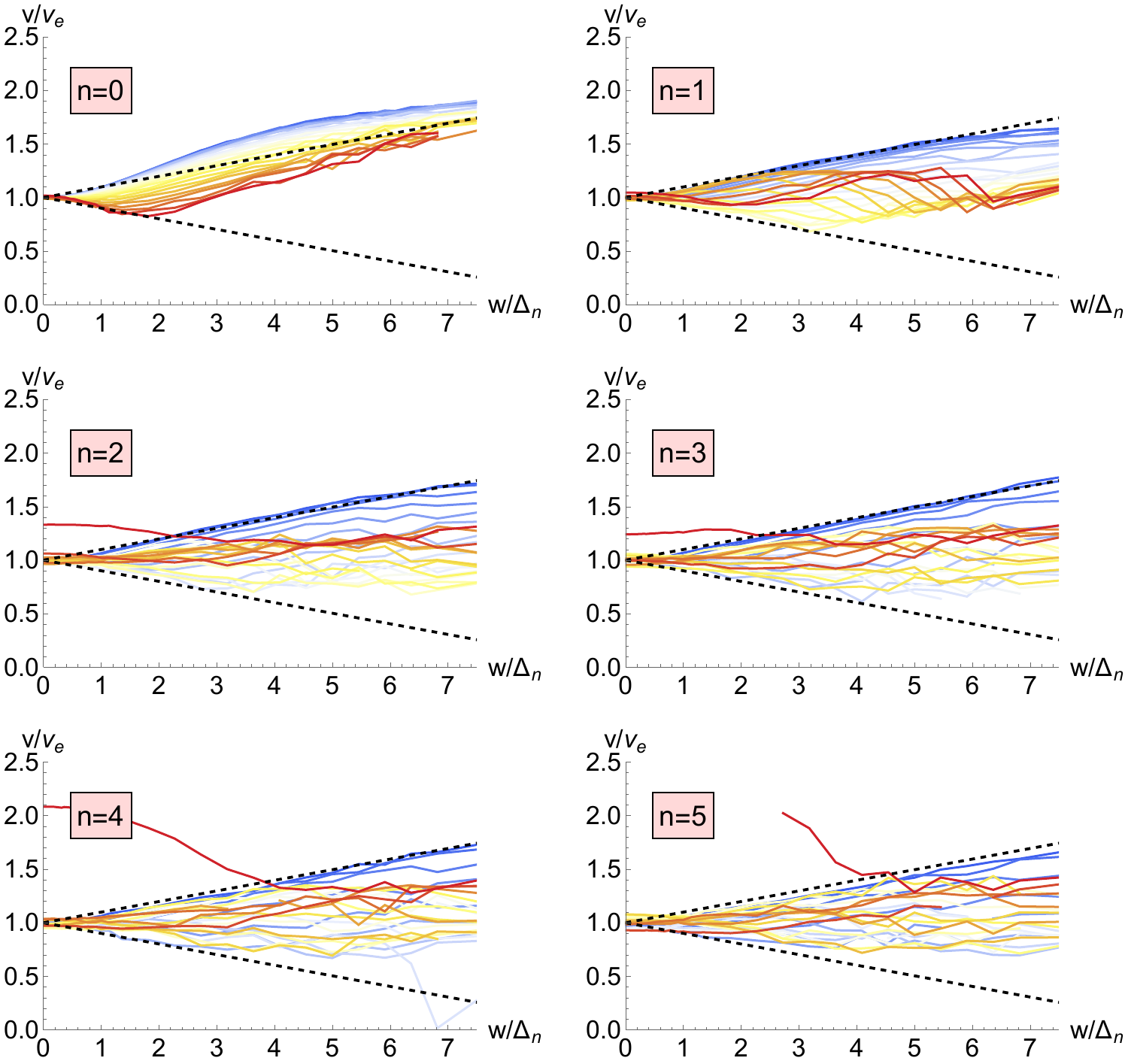}
		\caption{{\bf Velocity renormalization of the Chern insulator as function of winding number $n$} The color gradient represents different energies in the spectrum. The band center (blue) is renormalized more strongly than the band edge (red). The dark red curve is at an energy out of the optimization window and deviates strongly from the targeted velocity in the large $n$ cases. Up to disorder strength $w\approx 7$-fold multiples of the wrapped bandwidth $\Delta_n$, the modes are stable. For larger disorder, there is breakup since the lattice eigenstates no longer resemble plane waves. The velocity renormalization is strongly energy-dependent, non-monotonic and lies in a linear envelope. This chaotic behavior is caused by the energy-dependent contribution due to the bulk bands. }
		\label{fig:ren2a}
	\end{figure*}

	In Fig.~\ref{fig:ren2a}, the velocity renormalization of the wrapped edge as function of disorder strength $w$ and wrapping number $n$ is shown. The color gradient represents different positions in the spectrum. The band center (blue) is renormalized more strongly than the band edge (red). The dark red curve is at an energy out of the optimization window and deviates strongly from the targeted velocity in the large $n$ cases. Up to disorder strength $w\approx 7$-fold multiples of the wrapped bandwidth $\Delta_n$, the modes are stable. For larger disorder, there is breakup since the lattice eigenstates no longer resemble plane waves. The velocity renormalization is strongly energy-dependent, non-monotonic and lies in a linear envelope 
	\begin{align}
		v_{\rm env}(w) = v_e \pm 0.1 w / \Delta_n
	\end{align}
	(dashed line in panels of Fig.~\ref{fig:ren2a}). 
	The chaotic behavior inside the envelope is caused by the energy-dependent contribution due to the bulk bands.

	\section{Comparison to momentum space edge model}
	\label{sec:comp}
	In this appendix, we study a continuum model purely for the edges without the bulk states. First, we derive the self-energy correction leading to velocity renormalization analytically. Then we show that the velocity renormalization also occurs within a numerical model with perfectly linear dispersion.
	
	\subsection{Perturbative treatment: Self energy and velocity renormalization}
	\label{sec:self}
	
	We can understand this intuitively by looking at perturbative corrections to the self energy, $\Sigma_k(\omega)$, in the weak disorder limit~\cite{bruus2004many}:
	\begin{align}
		\Sigma_k(\omega) &= \sum_q \dfrac{S(q)}{\omega-\epsilon_{k+q}}.
	\end{align}
	Here,  $S(q)$ is the correlation function of the disorder in Fourier space and is equal to $S(q) = \frac{w^2}{12}$ for an uncorrelated identically distributed potential $V(x) \in [-w/2,w/2]$ in real space. Since there is no backscattering, this sum is purely real and we can convert it into a principal value integral:
	\begin{align}
		\Sigma_k(\omega) &= P.V. \int\dfrac{\mathrm{d} q}{2\pi} \dfrac{S(q)}{\omega-\epsilon_{k+q}}.
	\end{align}
	When we assume the white noise limit, $S$ is constant and the integral given by
	\begin{align}
		\Sigma_k(\omega) &= \dfrac{S}{2\pi v} \ln\left|\dfrac{\omega-\Delta}{\omega+\Delta}\right|, 
	\end{align}
	This expression is purely real and does not depend on $k$ but has an energy dependency for finite $\Delta$. We will therefore write $\Sigma(\omega)$ from now on. In the continuum limit $\Delta\rightarrow \infty$, the correction to the self-energy vanishes. 
	
	Therefore, the average Green's function can be written as:
	\begin{align}
		G_k(\epsilon) = \dfrac{1}{\epsilon-\epsilon_k -\Sigma_k(\epsilon)}
	\end{align}
	renormalizes the dispersion, allowing us to find the effective dispersion relation defined by $\tilde{\epsilon}_k$, by  setting the denominator of the average Green's function to zero:
	\begin{align}
		\tilde{\epsilon}_k-\epsilon_k -\Re \Sigma(\tilde{\epsilon}_k). &=0 \label{eq:spect}
	\end{align}
	For $\epsilon$ near $\tilde{\epsilon}_k$, we can linearize (i.e., perform a Taylor expansion of the self energy $\Sigma_k$):
	\begin{align}
		\epsilon-\epsilon_k -\Re\Sigma_k(\epsilon) = (\epsilon-\tilde{\epsilon}_k) (1-\partial_\epsilon \Re \Sigma(\epsilon)).
	\end{align}
	We may thus write  
	\begin{align}
		G_k(\epsilon) &=  \dfrac{Z(\epsilon)}{\epsilon-\tilde{\epsilon}_k}, 
		& Z(\epsilon)&\equiv (1-\partial_\epsilon \Re \Sigma(\epsilon))^{-1}.
	\end{align}
	for the average Green's function. 
	
	We now determine the renormalized spectrum by making an ansatz $\tilde{\epsilon}_k=\epsilon_k + \delta_k$, with $\delta_k$ small since we assume weak disorder. This means Eq.~\eqref{eq:spect} can be simplified, in the leading-order Taylor expansion, as:
	\begin{align}
		\delta_k - \Re \Sigma(\epsilon_k) &= \mathcal{O}(\delta_k^2)
	\end{align}
	This determines the renormalized spectrum:
	\begin{align}
		\tilde{\epsilon}_k = \epsilon_k + \Re \Sigma(\epsilon_k).
	\end{align}
	The renormalized group velocity therefore changes:
	\begin{align}
		\tilde{v}_k = v_k\left(1 + \partial_\epsilon \left. \Re \Sigma(\epsilon)\right|_{\epsilon=\epsilon_k}\right).
	\end{align}
	Using the quasiparticle weight $Z(\epsilon)$, we can express the renormalization of velocity as function of $\epsilon$:
	\begin{align}
		\tilde{v}(\epsilon) &= Z(\epsilon) v(\epsilon).
	\end{align}
	Hence, the velocity changes due to the renormalization of the quasiparticle weight $Z$. 
	
	\subsection{Numerics: exact diagonalization}
	\label{sec:mom}
	Ref.~\cite{Stacey_1982} introduces a method to avoid fermion doubling and model discretized versions of topological surface theories without the need of simulating bulk degrees of freedom.
	
	The chiral edge Hamiltonian:
	\begin{align}
		H(x,x') &= \pi\left[1+i\cot\left(\pi\dfrac{ x-x'}{L}\right)\right] + V(x)\delta_{x,x'} \label{eq:ham_lat}
	\end{align}
	is nonlocal (has long-range hoppings decaying as $(x-x')^{-1}$) and has a perfectly linear dispersion within its finite bandwidth. This nonlocality translates to the discontinuity of the dispersion in momentum space, i.e., there is a jump at the Brillouin zone boundary.

\begin{figure}
		\centering
		
		\includegraphics[width=0.45\textwidth]{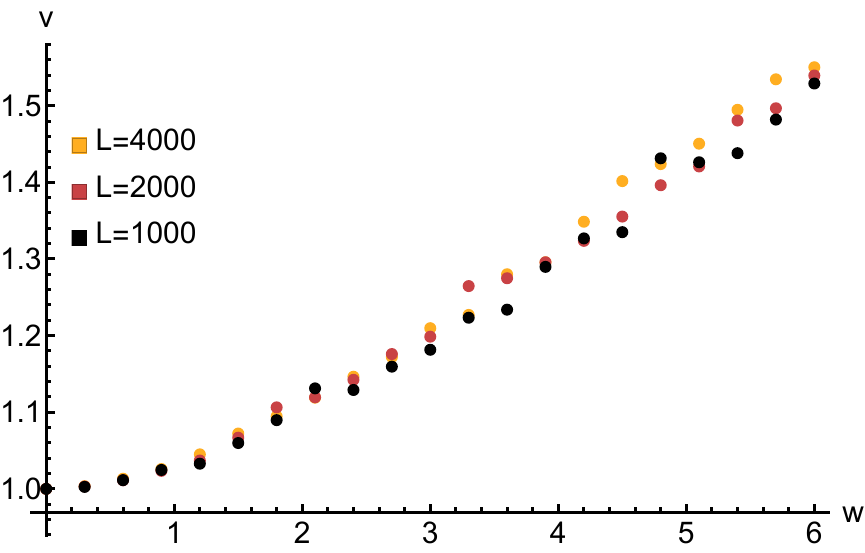}
		\caption {{\bf Continuum edge model velocity renormalization} Average velocity as function of disorder in the linear dispersion lattice model \eqref{eq:ham_lat}.  In the strong disorder regime, $w\gtrsim 1$, there is a linear relationship between the renormalized velocity and disorder. 
		}
		\label{fig:ren1a}
	\end{figure}

	In Fig.~\ref{fig:ren1a}, we show the renormalized velocity. For weak disorder, the dependence of $\delta v/v_e$ on disorder strength $w$ is quadratic. In the strong disorder regime $w\gtrsim 1$, the relationship between renormalization and disorder strength $w$ becomes approximately linear. These features (crossover quadratic into linear dependence) qualitatively resembles the phenomenology of the continuum edge Hamiltonian or full Chern insulator bulk model.

	\section{Momentum space toy model}
	\label{sec:scattering}
	Here we describe a discretized one-dimensional momentum-space model of the edge dispersion with realistic impurity matrix elements constructed to match those of the realistic fully two-dimensional Chern insulator.
	
	\subsection{Modelling allowed scattering processes}
	\label{sec:mod_imp}
	In the case where there is just one channel ($n=0$ extra wrappings), the impurity potential is a function of the of the momentum transfer $k-k'$. For higher windings, it can also depend on the branches explicitly. This can be seen by considering the matrix element due to one impurity $u_{{\bf r}_0}(x,y)$ localized around $x_0$ near the edge ($y_0\lesssim (2n+1)a$). Let $\psi_k ({\bf r})$ be the eigenstates of the full clean edge Hamiltonian and consider the matrix element with the impurity:
	\begin{align}
		\langle \psi_k| u |\psi_k'\rangle &= \int d^2{\bf r}\; \psi_k^*({\bf r})u_{{\bf r}_0}({\bf r}) \psi_{k'}(x,y)
	\end{align}
	From the exact diagonalization results in Ref.~\cite{Guglielmon2019}, we know that for $k= G + (k\mod 2\pi)$ in the $G$-th branch, the eigenstate $\psi_k$ is localized in the transverse direction near $y\sim G$. In the linear direction, it behaves like a plane wave $\psi_k(x,y\sim G)\sim e^{i k x}$. This gives the matrix elements an explicit dependence on the branch numbers $G, G'$.
	
	\begin{figure}
		\centering
		\includegraphics[width=0.32\textwidth]{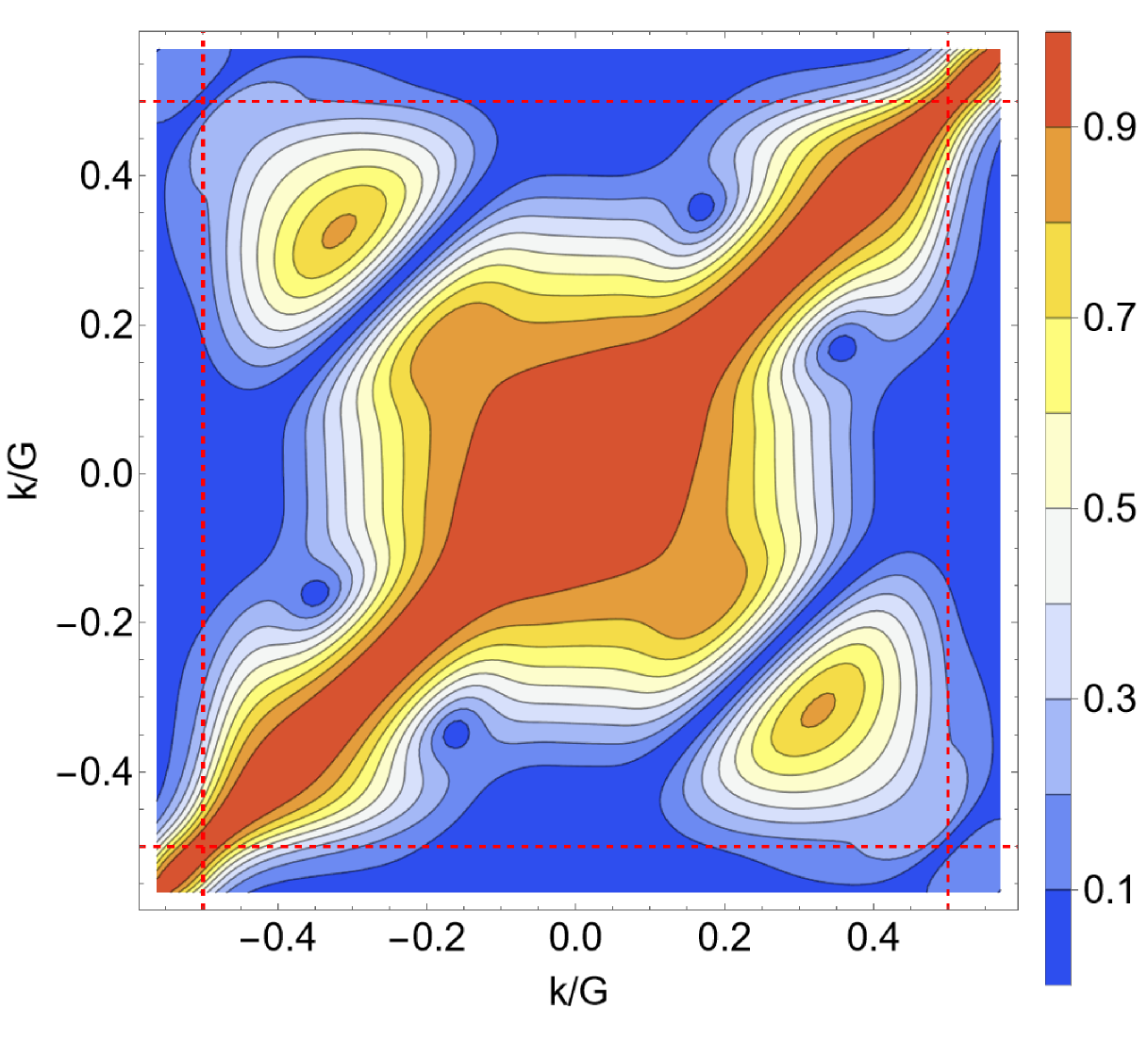}
		\includegraphics[width=0.32\textwidth]{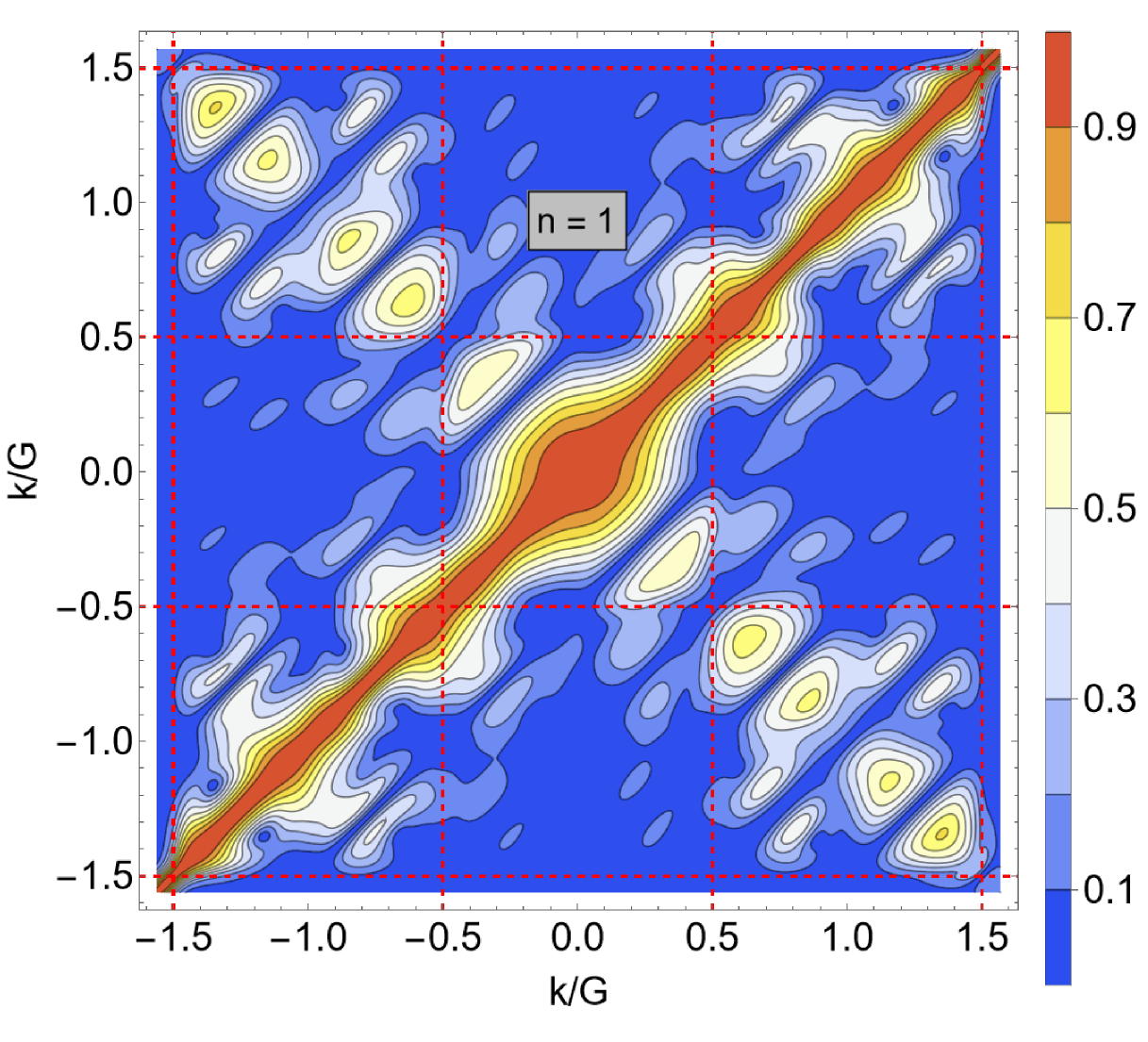}
		\includegraphics[width=0.32\textwidth]{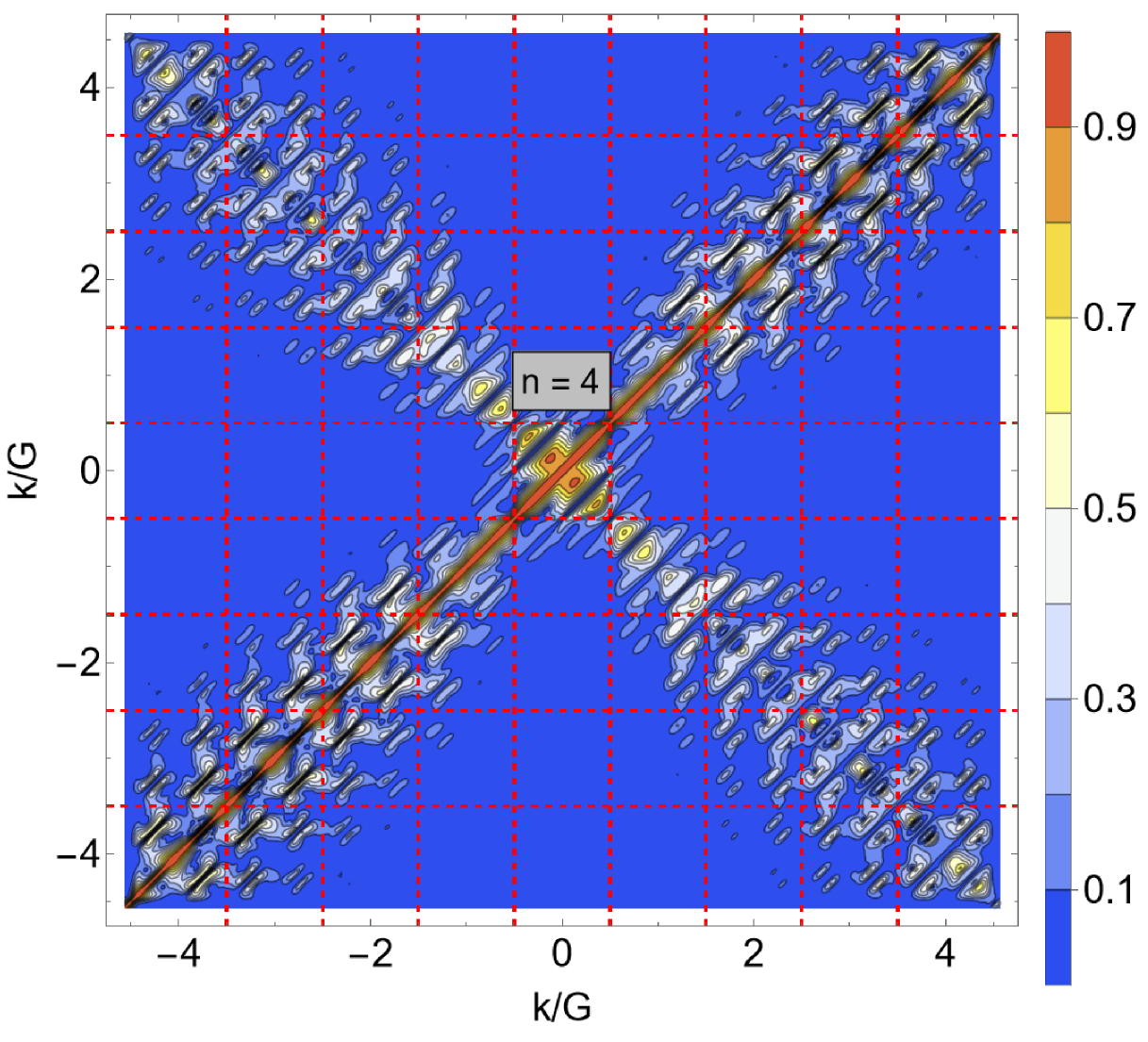}
		\caption{{\bf Real space numerics investigating impurity matrix elements in the optimized edge model} We study an impurity localized at the boundary and take the matrix elements $\left| \int d^2 {\bf r} u_{{\bf r}_0}({\bf r}) \psi_{k}({\bf r})^* \psi_{k'}({\bf r})\right| $ with clean states $k,k'$ (strength of this matrix element is shown in colors). An average over the impurity position ${\bf r}_0$ is taken. The Brillouin zone boundaries (extended zone scheme) are shown as dashed red lines. For $n=0$ (left panel), every state is talking to every other. In the middle and right panel, we show the more complicated $n=1,4$ cases with three or nine minibands. Different minibands are coupled only weakly by adding impurities. }
		\label{fig:imp_matrix2}
	\end{figure}

	\begin{figure}
		\centering
		\includegraphics[width=0.5\textwidth]{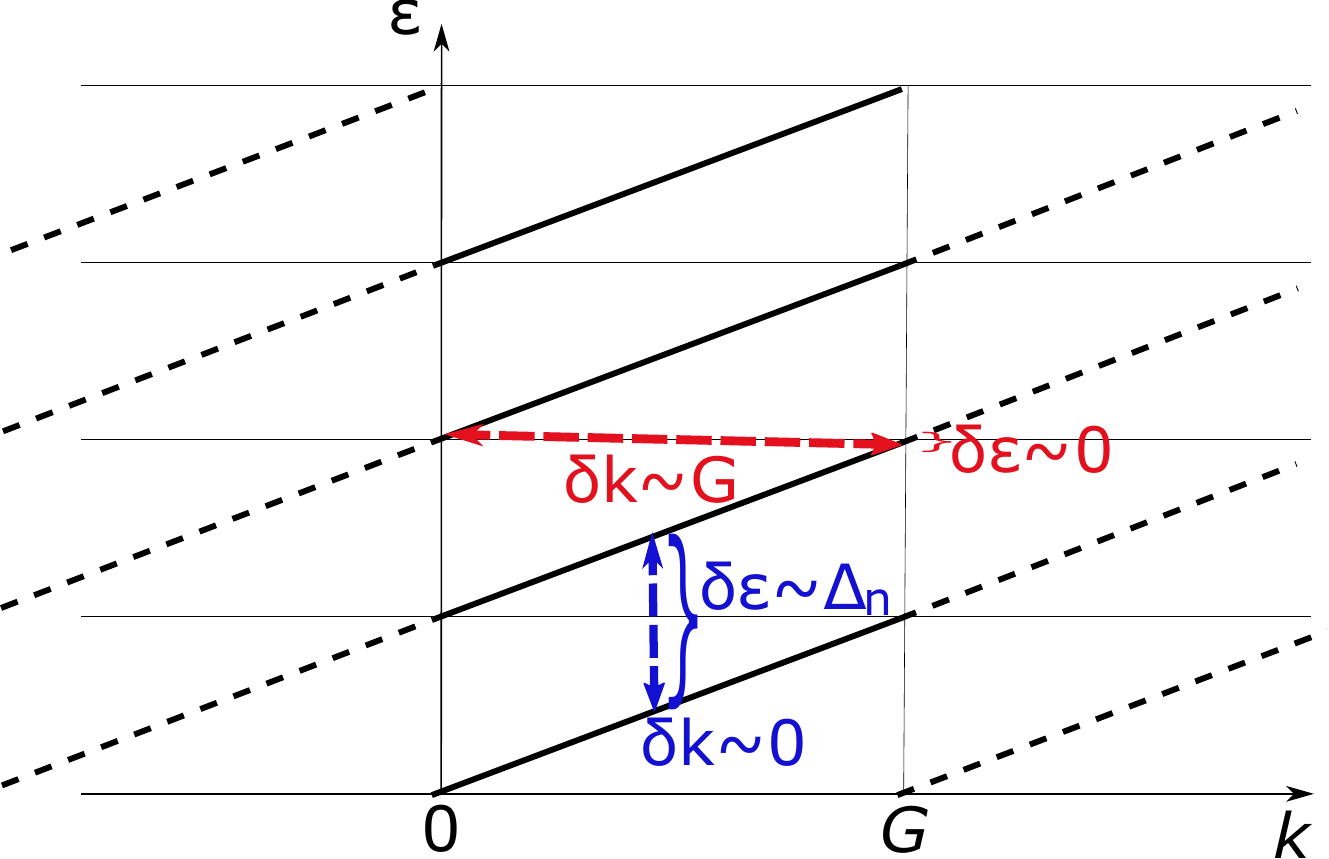}
		\includegraphics[width=0.35\textwidth]{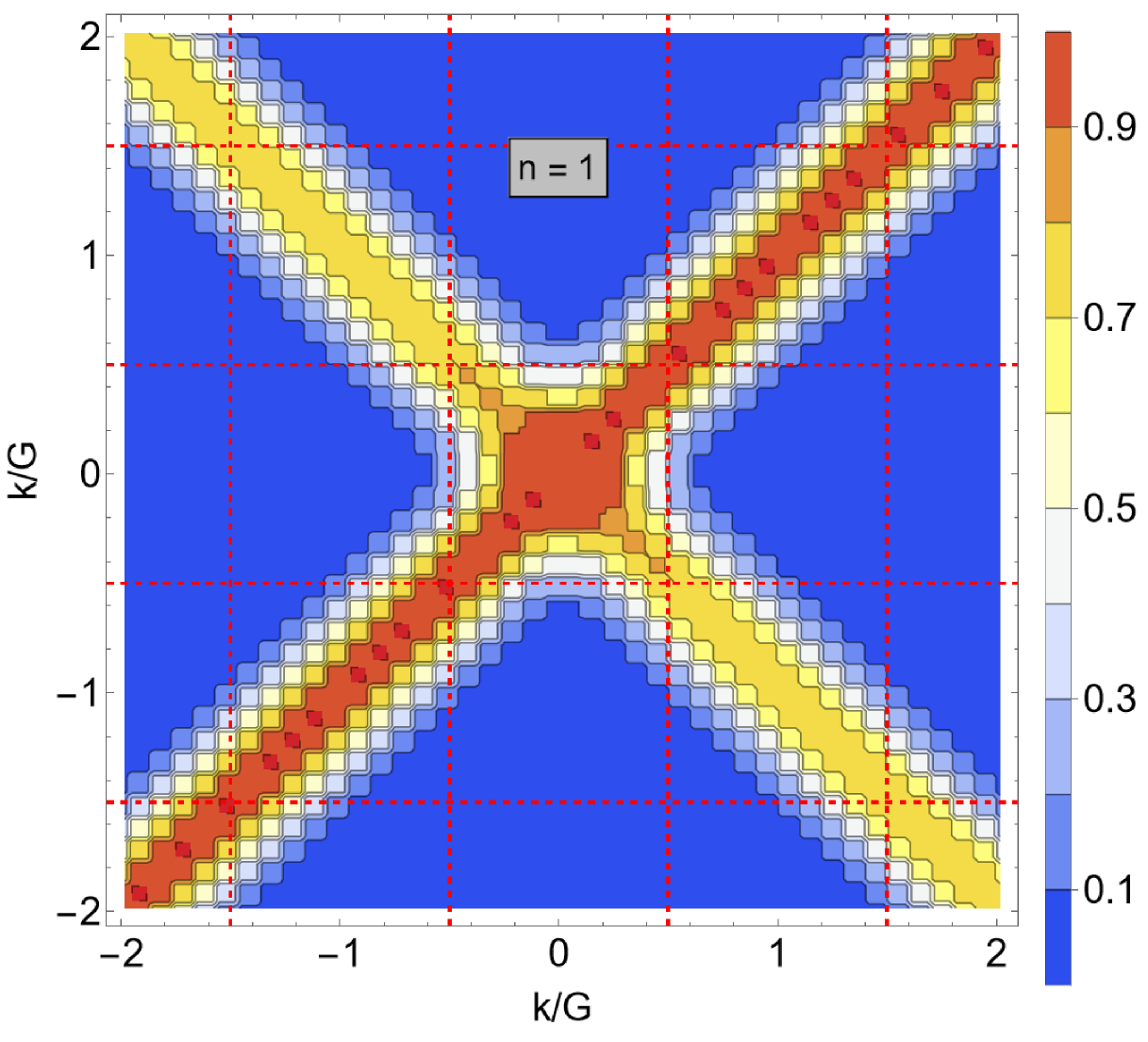}
		\caption{{\bf Dispersion and modelling of scattering} (a) Sketch of the dispersion of Hamiltonian~\eqref{eq:ham_low}. Two characteristic scattering processes changing by one band are shown as arrows. The process shown with the red arrow is only possible for initial $k_i\sim 0$ and final $k\sim G$ near the boundaries of the Brillouin zone. The energy difference $\delta \epsilon$ can be made small. Direct changes of the branch with arbitrary initial $k_i$ and final $k_f$ are strongly suppressed by multiples of the energy detuning $\Delta_n$ separating the branches (blue arrow). (b) Modelled matrix element capturing the features of the full bulk Chern insulator (data shown in Fig.~\ref{fig:imp_matrix2}) closely. }
		\label{fig:imp_matrix}
	\end{figure}

	In Fig.~\ref{fig:imp_matrix}, we show a numerical study of the localized impurity matrix elements $v_{k,k'}$, that confirms this assumption. In particular for higher winding $n>0$, the matrix elements are no longer a single-valued function of $k-k'$, but rather take a complicated dependency on both $k$ and $k'$. This implies that (a) intra-branch scattering can be approximately descibed by a pure potential $V(x)$ and (b) inter-branch scattering only weakly couples minibands $i$ and $-i$.
	
	With the data of Fig.~\ref{fig:imp_matrix2}, we justify the toy momentum space edge model in the main text. The toy model impurity matrix elements shown in panel (b) of Fig.~\ref{fig:imp_matrix} are chosen to closely resemble the ones of the bulk model.

	\subsection{Breakup length: numerical simulations}
	\label{sec:mom_break}
	We study the chiral edge state Hamiltonian~\eqref{eq:ham_low} in the main text by discretizing momentum space~\cite{Chou2014chalker} to find the velocity renormalization. Here, we also compute the breakup length and compare expectations from Eq.~\eqref{eq:prop} and propagation simulations (lower panels of Fig.~\ref{fig:ren5}). 
	We include realistic impurity matrix elements comparable with the bulk model as motivated in Sec.~\ref{sec:scattering}.
	
	In order to test the effect of the disorder in our effective model, we prepare a gaussian wavepacket with energy bandwidth $\delta\epsilon$. The matrix exponential required for time evolution is dense, yet numerical calculations remain efficient for moderate system sizes and winding numbers. Our observables are position $x(t)$ and width of the wavepacket $\sigma_x(t)$ in real space throughout time evolution. From these, the velocity $v(t) = \dot{x}(t)$ is derived. Using the results for $\sigma_x(t)$, we can assign a breakup time $\tau$ by defining this the to be the time when the real space width has reached a constant, $C$ times the original width $\sigma_x(t=\tau)=C\sigma_x(t=0)$. The results depend only weakly on the coefficient $C>1$, we choose $C=4.5$ throughout.
	
\begin{figure}
		\centering
		\includegraphics[width=0.99\linewidth]{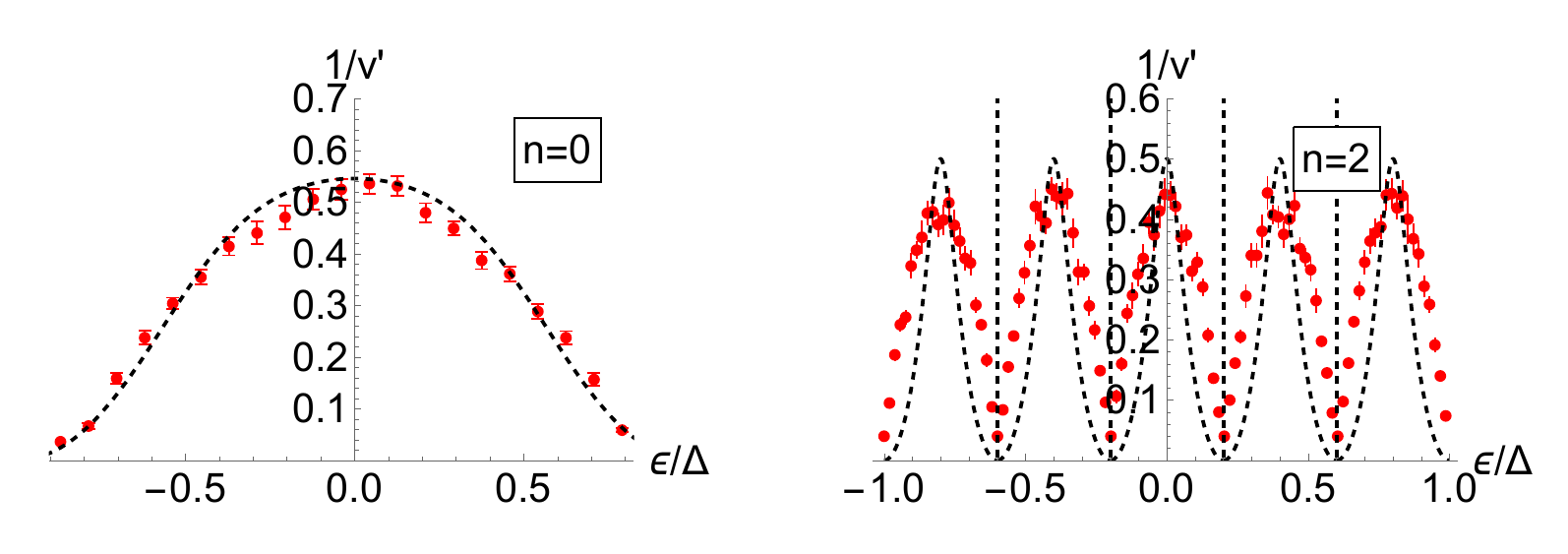}
		\caption{{\bf Velocity renormalization and breakup:} (a) Velocity renormalization, $\delta v$ for $n=0$ and (b) $n=2$.  The black-dashed lines are the analytical result from Eq.~\eqref{eq:prop}, and the red dots are breakup times from numerical time evolution of the continuum edge model.  The system size is $L=2500$, disorder strength $w=0.9\Delta_n$ and $\delta \epsilon = \Delta_n/10$. 
		} 
		\label{fig:ren5s}
	\end{figure}

	We confirm our prediction Eq.~\eqref{eq:vel_ren_ana} in the main text for velocity renormalization in this system. Here, we tend to the wave packet breakup induced by this effect (shown in Fig.~\ref{fig:ren5s}). Near the band center, the velocity is almost not renormalized. Therefore, $1/v'$ is ill-defined for very narrow wavepackets and smoothening over the energy bandwidth $\delta\epsilon$ is absolutely necessary to obtain a meaningful finite result. This is the cause of the large the error bars near the band centers in Fig.~\ref{fig:ren5s}. Since the subbands are only weakly coupled in the bulk model (see discussion in Sec.~\ref{sec:mod_imp}), there are peaks of the breakup length in each subband center. \jk{Near the band edges the (gaussian in energy) wavepacket has overlap into adjacent minibands. Our simple treatment does not take this case into account. A way to better control our approximation would be to choose a narrower wavepacket, this has the computational downside, that longer devices are necessary to study breakup. }
	
	\section{Robustness of edge states at strong disorder}
	\label{sec:robust}
	The main text chiefly considered the case of weak disorder tractable with perturbative methods. In order to go beyond this, we have used the kernel polynomial method (KPM) in order to investigate the fate of the edge states at intermediate and strong disorder. Since we modify only the edges of the system to achieve multiple winding, the bulk gap is expected to be robust in a semi-infinite slab. However, for large windings $n$ a significant fraction of the system is engineered in the finite width systems considered here. We choose strips of width $W=75$ and length  $L=400$.

\begin{figure}
		\centering
		\includegraphics[width=0.9\textwidth]{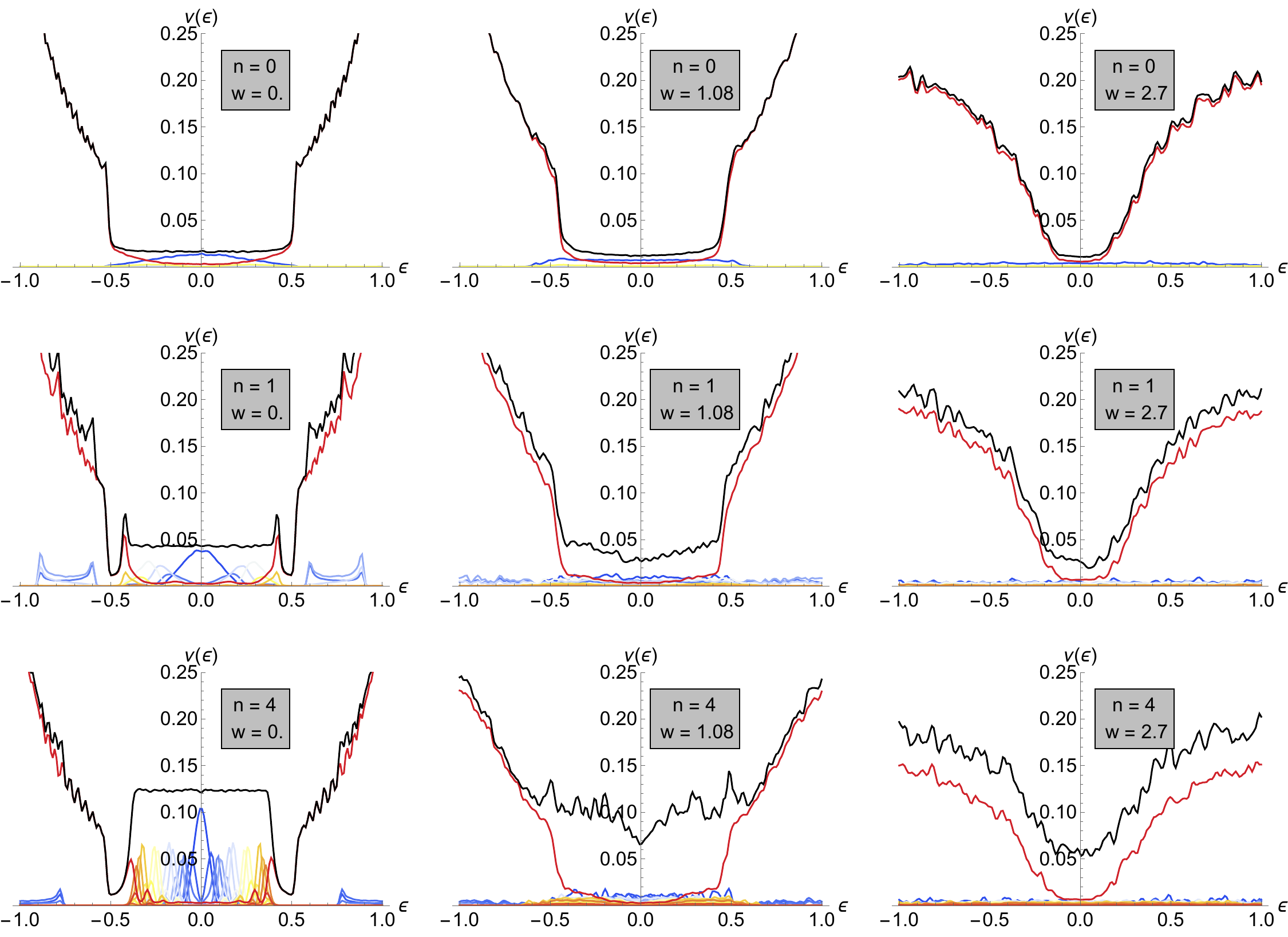}
		\caption{{\bf Strong disorder study of the edge states with winding} Using KPM, we compute the transverse LDOS from Eq.~\eqref{eq:dos} for winding numbers $n=0,1,4$ and disorder strengths $w=0, 1.08, 2.7$. The global DOS $\nu(\epsilon)$ is shown as solid black line. For different $y$, the LDOS $\nu(y,\epsilon)$ is shown color gradient coded: blue is $y=0$ at the edge turning over yellow into red further away from it. The bulk LDOS (summed over all $y>3n$) is shown in red.}
		\label{fig:dos}
	\end{figure}

	The disorder strength $w$ is measured in units of the hopping here. In these units, the bulk gap is $\Delta_0 = 0.5$ and the minigaps $\Delta_n \sim 0.375/(2n+1)$ (a fraction of $\sim 70-75\%$ of states in the gap can be optimized well to a fixed common velocity). Our observable is the local density of states (LDOS) 
	\begin{align}
		\nu(\epsilon, y) &\equiv \int_0^{L} dx\;\nu(\epsilon, x, y), & \nu(\epsilon, x, y)&=\sum_n |\psi_n(x, y)|^2\delta(\epsilon-E_n) \label{eq:dos}
	\end{align}
	summed over the transport direction $x$. For $y$ near the edge, $y=0$, this is a direct measure of the number of edge states. The global density of states (DOS) $\nu(\epsilon)$ is obtained by summing $\nu(\epsilon, y)$ over all $y$.
	The clean systems are shown in the left column of Fig.~\ref{fig:dos}. We can nicely observe the band center edge states to be localized right at the boundary of the system. For higher and lower minibands the states have support further away from the boundary.
	
	The bulk gap does not close even in the presence of $W\gtrsim \Delta_0$ (middle column). Only for $W> 5\Delta_0$ the amount of edge states significantly drops (right column).
	For higher winding, the velocity is lower resulting in larger DOS in the gap. The DOS in the optimized region (75 \% of the gap) is constant in the clean systems. Signatures of the velocity renormalization discussed in the main text are deviations from uniform DOS visible in the middle column of Fig.~\ref{fig:dos}. Even at strong disorder the larger DOS due to the slower average velocity is still visible.
	
\end{document}